\documentclass[onecolumn, a4paper]{article}
\usepackage[left=3.5cm,right=3cm,top=3cm,bottom=3cm]{geometry}
\usepackage{authblk}
\usepackage[hidelinks]{hyperref}
\hypersetup{
  colorlinks   = true, 
  urlcolor     = black, 
  linkcolor    = red, 
  citecolor   = green 
}

\usepackage{array}
\usepackage{rotating}
\usepackage{graphicx}%
\usepackage{multirow}%
\usepackage{amsmath,amssymb,amsfonts}
\usepackage{amsthm}%
\usepackage{booktabs}%

\title{Disease Classification and Impact of Pretrained Deep Convolution Neural Networks on Diverse Medical Imaging Datasets across Imaging Modalities}

\author{Jutika Borah$^{1}$}
\author{Kumaresh Sarmah$^{1}$}
\author{Hidam Kumarjit$^{1}$}

\affil{$^1$Gauhati University, Guwahati 781014, India
\\
\texttt{\href{mailto:borah\_jutika@gauhati.ac.in}{borah\_jutika@gauhati.ac.in}  \hspace{0.3cm} \href{mailto:kumaresh@gauhati.ac.in}{kumaresh@gauhati.ac.in}  \href{mailto:kumarjit\_hidam@gauhati.ac.in}{kumarjit\_hidam@gauhati.ac.in}
}
}

\date{}

\begin{document}
\maketitle

\begin{abstract}
    Imaging techniques such as Chest X-rays, whole slide images, and optical coherence tomography serve as the initial screening and detection for a wide variety of medical pulmonary and ophthalmic conditions respectively. This paper investigates the intricacies of using pretrained deep convolutional neural networks with transfer learning across diverse medical imaging datasets with varying modalities for binary and multiclass classification.  We conducted a comprehensive performance analysis with ten network architectures and model families each with pretraining and random initialization. Our finding showed that the use of pretrained models as fixed feature extractors yields poor performance irrespective of the datasets. Contrary, histopathology microscopy whole slide images have better performance.  It is also found that deeper and more complex architectures did not necessarily result in the best performance. This observation implies that the improvements in ImageNet are not parallel to the medical imaging tasks. Within a medical domain, the performance of the network architectures varies within model families with shifts in datasets. This indicates that the performance of models within a specific modality may not be conclusive for another modality within the same domain. This study provides a deeper understanding of the applications of deep learning techniques in medical imaging and highlights the impact of pretrained networks across different medical imaging datasets under five different experimental settings.
\end{abstract}

\section{Introduction}\label{sec1}

Clinical decision-making through diagnosis is a complicated process. Diagnostic processes usually encompass visual inspection of digital images and interpretation of the findings as per established protocols \cite{bib1}. In medical imaging, disease diagnosis is a critical, and time-consuming process \cite{bib2, bib3}, which relies on the knowledge, experience, and reasoning skills of an expert. Chest X-rays (CXRs), Optical Coherence Tomography (OCT), and Whole Slide Images (WSIs) are commonly investigated for diagnosis of thoracic and pulmonary diseases \cite{bib4, bib5, bib6}, ophthalmic diseases respectively. Recently, deep learning (DL) has shown the potential to automate the process of medical image interpretation for faster and more accurate healthcare delivery \cite{bib7, bib8, bib9, bib10, bib11}. However, DL holds a few challenges such as it requires a huge amount of labeled training data for training in a supervised manner, whereas finding labeled data in the medical domain is hard. These challenges have prompted researchers to leverage the benefits of transfer learning (TL) in medical imaging as it can generalize and transfer information from prior experience to new conditions. DL typically suffers from insufficient data distribution, shifts during test time, and computation power. TL deals with these by utilizing the already learned knowledge without training a model from scratch. However, TL performance varies when distribution shifts occur between the source and target task which may affect the model’s performance in prospective clinical settings. However, little study has been done in medical imaging as of now on the truth, reliability, and efficacy of TL across new distribution shifts with the change of imaging modality within the same domain \cite{bib12, bib13}. 

To date although many neural networks have been developed for the classification of CXR, OCT, and WSI, however, the extent to which pretraining on ImageNet improves performance on diverse medical image classification tasks such as CXR, OCTs, and WSIs is not fully understood. Some studies have suggested that the benefits of TL may be limited due to the fundamental differences in data characteristics and task specifications between natural images and medical images. The authors in \cite{bib14} explored TL with ResNet50 and InceptionV3 on retinal fundus images and, the CheXpert dataset. Their study concluded that through pretraining there is little performance improvement. The research work reported in \cite{bib15, bib16} used pretrained networks on CheXpert CXR datasets, Covid19 datasets, and retinal fundus images respectively. The authors in\cite{bib18} studied the performance of sixteen pretrained models on twelve natural image datasets. They found a correlation between the accuracies of the networks on using pretrained networks as fine-tuning or fixed feature extractors. They also found that, with models’ architectures, the performance improved across datasets. In our study, using the models as fixed feature extractors on CXR, and OCT classification gave poor accuracies and area-under-the-curve (AUC). However, on the histopathology WSI dataset, the models show a better performance compared to CXR and OCT. With these exceptions, our approach differs from that of the cited works in that we considered three independent datasets belonging to different imaging modalities. 

To investigate these questions, we conducted an empirical study to examine the impact of pretraining on training from scratch with three independent diverse datasets of varying modalities. We explored the importance of factors such as fixed ImageNet weights on medical imaging datasets, the size of the model, and a varying number of classes (binary vs. multiclass). Here, we used ten different DCNNs with different architectures VGG (16, 19), ResNet (50, 101, 152), InceptionV3, InceptionResNetV2, Xception and DenseNet (121, 201) for supervised medical image classification for both binary and multiclass. We trained and evaluated the performance of these models on publicly available datasets both with pretraining and random initialization. This investigation and findings can be used to validate other medical imaging modalities. Our findings suggest that pretraining on a large, general-purpose natural ImageNet dataset can provide significant benefits for medical classification tasks when there is a substantial domain mismatch. However, the extent of improvement varies depending on specific medical imaging tasks, dataset characteristics, and information. In general, we found that the benefits of pretraining were less when the scale and properties of the data were diversified. Thus we observed that the same correlation between networks may not be necessarily consistent as we shift data from one modality to another within the same domain. This highlights the importance of considering the effects of efficient network architectures with varying modalities for real-time clinical scenarios. This study aims to contribute to the ongoing research on the use of DCNNs with TL for medical image classification and interpretation. The findings could help future studies and the development of DL-based diagnostic tools for medical image diagnosis. 

The main contributions of the paper are as follows:
\begin{itemize}
    \item For models with pretraining and random initialization, we observed that there is no relationship between the performance on ImageNet and the performance on medical imaging datasets of varying modalities. This finding suggests that improvements in ImageNet are not analogous to medical imaging tasks with architectural enhancements. 

    \item With ImageNet pertaining, performance improvement is observed across the model architectures, but a higher boost in performance is obtained in the models with smaller architecture for different datasets of varying modality (CXR, OCT, and WSI).

    \item If training only a few top layers of the networks whether with pertaining or random initialization suggests that the models can perform better with decreased parameters on medical imaging tasks without a significant drop in performance. In terms of architectural design, this finding indicates that lighter architectural design, with reduced model size, can be effective.  
    
\end{itemize}

\section{Materials and Methodology}\label{sec2}
This section gives brief details of the experimental study design of the proposed work.
\subsection{Datasets}\label{subsec1}
\begin{itemize}
    \item \textit{Pneumonia CXR} \cite{bib19}: A total of 5,863 CXR images are present in the dataset grouped into Normal and Pneumonia classes. The dataset has three folders split into training, validation, and test sets. Class Imbalance exists in the dataset with more numbers of pneumonia images compared to normal images. The training set has a total of 5,247 images of which 3,906 images belong to pneumonia and 1,341 images of normal subjects. We grouped the images into separate training and validation datasets while keeping a set separately for testing with 234 normal images and 390 pneumonia images. The training set is now comprised of 4000 samples and the validation set is comprised of 1043 samples. These CXR datasets were collected from retrospective cohorts of pediatric patients from Guangzhou Women and Children’s Medical Center, Guangzhou.

    \item \textit{OCT} \cite{bib19}: This dataset contained 1,09,309 OCT images with separate folders for training and test sets for independent patients. Each folder has four classes of images, viz: choroidal
    neovascularization (CNV), diabetic macular edema (DME), DRUSEN, and NORMAL. The training set contained 1,08,309 OCT images (37,205 CNV, 11,348 DME, 8,616 DRUSEN, and 51,140 NORMAL), whereas the test set had 1,000 images (250 images for each class). The structure of the dataset and labeling of the images were given in order of disease type, randomized patient ID, and image number of the patient. A validation set was not available in the original dataset. So, we have generated a validation set by splitting the training set in the ratio of 80\%-20\%. Thus, our modified dataset contains 86647, 21662, and 1000 images in training, validation, and test sets respectively.

    \item \textit{LC25000} \cite{bib20}: LC25000 is a histopathology image pathology with 25,000 color images in 5 classes. The 5 classes are divided into separate subfolder folders each containing 5,000 images of histologic entities namely, lung adenocarcinoma, lung benign tissue, lung squamous cell carcinoma, colon adenocarcinoma, and benign colonic tissue. The images are publicly available and are de-identified, HIPAA compliant, and validated. The images were of sizes of 768 x 768 pixels in jpeg file format. For this experiment, we have taken only lung carcinoma cases and discarded the colon adenocarcinoma cases. The dataset has been split into train, validation, and test sets. The training set contains 9600 histopathology lung adenocarcinoma samples, 1680 samples in the validation set, and 720 samples in the test set. 
    
\end{itemize}

\begin{figure}[ht]
\centering
\includegraphics[width=1\textwidth]{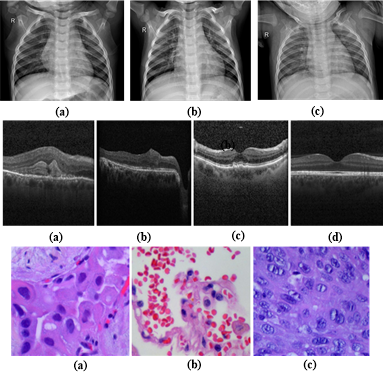}
\caption{Image from top (a) Normal CXR, (b) Bacterial Pneumonia CXR, and (c) Viral Pneumonia CXR. Image at the middle (a) Choroidal neovascularization (CNV), (b) Diabetic macular edema (DME), (c) Drusen, and (d) Normal. Image at the top, WSI Histopathology (a) Lung adenocarcinoma, (b) Lung Normal, and (c) Lung squamous cell carcinoma.}\label{fig1}
\end{figure}

\begin{table}[ht]
\centering
\caption{Architectural Description of Pretrained DCNN Models}\label{tab1}
\begin{tabular}
{@{}llll@{}}
\toprule
Pretrained Models & Parameters      & log values & ImageNet \\
                  & (in millions)   & (**)       & Top-5\% accuracy\\   
\midrule
VGG16    & 138.4 & 8.14 & 90.1   \\
VGG19    & 143.7 &  8.2 & 90.0    \\
ResNet-50    & 25.6 & 7.4 & 92.1  \\
ResNet-101    & 44.7 &  7.6  & 92.8  \\
ResNet-152    & 60.4 &  7.8  & 93.1  \\
Inceptionv3    & 23.9 & 7.3  & 93.7  \\
Inception-ResNet-v2    & 55.9 & 7.7 & 95.3 \\
Xception    & 22.9 & 7.3  & 94.5   \\
DenseNet121    & 8.1 & 6.8 & 92.3 \\
DenseNet201    & 20.2 & 7.2  & 93.6 \\
\bottomrule
\end{tabular}\\
\footnotesize ** indicates parameters of models in logarithmic scale (base 10). The logarithmic scale is chosen for convenience in analyzing the parameters of the models that are in millions.
\end{table}

\subsection{Data Preparation and Preprocessing}\label{subsec2}
In the pre-processing stage, the images in the datasets are resized as the images are of varying sizes due to computational constraints as it would be difficult for the networks to train optimally. And so, resizing ensure that all the images are under the same dimension and that these can be fed into the neural network for training. In a nutshell, huge file sizes are computationally expensive, requiring larger power and memory sizes. Here, we resized the images of both datasets to 256 × 256 × 3 with no aspect ratio preservation. Then, the pixels of all the resized images are normalized to [0, 1]. A few examples from the three datasets are shown in Figure~\ref{fig1}.

\subsection{Model Architectures}\label{subsec3}
We used ten widely used pretrained DCNN architectures: VGG (16, 19), ResNet (50, 101, 152), Inceptionv3, Inception-ResNet-v2, Xception, and DenseNet (121, 201). To assess the effectiveness of these models on the two different imaging datasets for binary and multiclass classification tasks. To obtain a comprehensive understanding of the model’s performance, we experimented with two different initialization approaches pretraining and random initialization. In pretraining existing pretrained ImageNet weights from large-scale image recognition tasks were taken, while for random initialization randomly initialized weights were taken. Table~\ref{tab1} shows all the pretrained models with their respective parameters in millions and ImageNet top 5\% accuracy.

\begin{figure}[ht]
\centering
\includegraphics[width=1.0\textwidth]{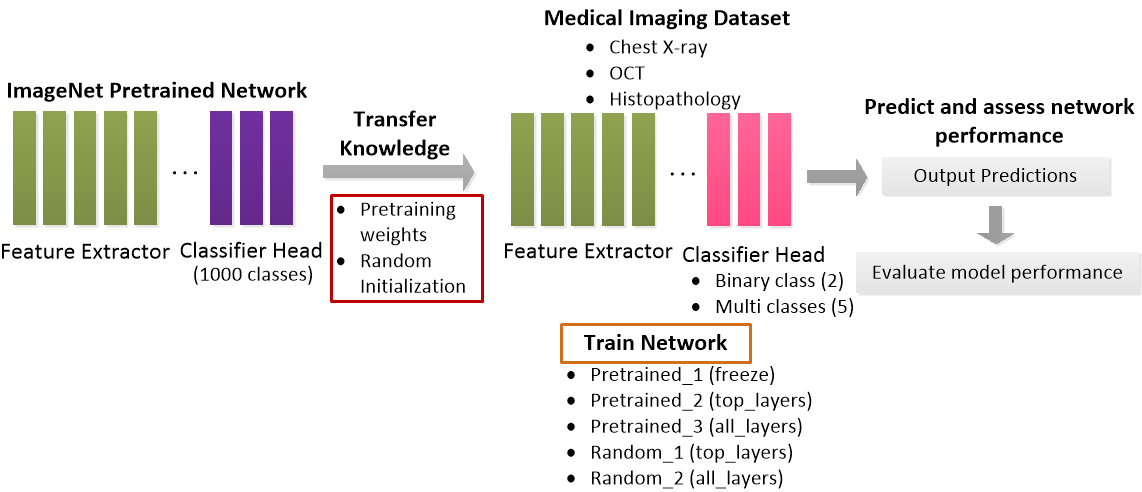}
\caption{Block representation of experimental design and evaluation framework. The figure shows all five different settings for performance evaluations with pretraining and random initializations.}\label{fig2}
\end{figure}

\section{Experimental Settings, Training Strategies and Performance Evaluation}\label{sec3}
We designed an experimental framework for training and evaluation of the performance of different DCNN models on CXR, OCT, and WSI datasets for automatic classification and interpretation of diseases. Figure~\ref{fig3} shows a visual representation of the overview of the different experimental and evaluation settings under study. 

\emph{\textbf{A. Settings with pretraining:}}
\begin{itemize}
    \item Pretrained\_1(freeze): Here, we keep all the CNN layers as fixed feature extractors and fine-tune them with our chest X-ray dataset. This is done for all the ten pretrained models with their corresponding pretrained ImageNet weights.

    \item Pretrained\_2(top\_layers): In this setting, we fine-tuned the top layers of the networks and trained them on our chest X-ray dataset while freezing the bottom layers. Here, training for all the pretrained models is accomplished with their respective pretrained weights.

    \item Pretrained\_3(all\_layers): In this setting, we retrain all the layers of the ten pretrained models with their corresponding pretrained ImageNet weights on our chest X-ray image data.
    
\end{itemize}        

\emph{\textbf{B. Settings with random initialization:}}

\begin{itemize}
 
    \item Random\_1(top\_layers): In this setting, we fine-tuned the top layers of the networks and trained them on our chest X-ray dataset while freezing the bottom layers. Here, training for all the pretrained models is accomplished with random weights initialization during training. 

    \item Random\_2(all\_ layers): In this setting, we fine-tuned the top layers of the networks and trained them on our chest X-ray dataset while freezing the bottom layers. Here, training for all the pretrained models is accomplished with their respective pretrained weights.
            
\end{itemize}

For training purposes, we fine-tuned the models and used their pre-trained weights for supervised binary and multi-label classification tasks. The models are trained using Keras deep learning library with TensorFlow backend, Adam optimizer, and cross-entropy loss function is used across the whole training process. We trained each model for 300 epochs under each setting. For each experimental setting, we train the models 3 times each. The different hyperparameters are chosen: learning rate (1e-3), and batch size (16). The learning rate decayed by a factor of 0.2 after 5 epochs if the networks’ validation loss plateaus (stopped improving). Training these deep architectural models on larger batch sizes requires larger memory and high-performance hardware. So, to compensate we took smaller batch sizes of 16 for this study. The whole training, testing, and evaluation are performed on NVIDIA GeForce RTX 3060Ti with CUDA 10 and Intel(R) Core (TM) i7-11700F running on Windows 11.

Here, we did not work on optimization of any specific network for best model selection, rather we focused more on their performances and evaluations. For the whole training process, all the hyperparameters were kept constant. For comparisons and evaluations of performance, we emphasized standard evaluation metrics like Area Under the ROC-Curve (AUC), precision, sensitivity, and specificity \cite{bib21}. A comprehensive performance analysis was conducted to compare the results obtained with different DCNN architectures and model families.

\section{Experimental Results and Analysis}\label{sec4}
Figure~\ref{fig3}, shows different experimental settings and study designs of our work. The experimental results for average CXR AUCs, OCT AUCs, and WSI AUCs each with pretraining and random initializations are visually depicted through scatter plots in Figure~\ref{fig3} (a), (b), and (c). The scatter plot clearly illustrates the performance difference in AUCs relative to the number of parameters of the models representing their complexity (refer Table~\ref{tab1}). This comparison is made across five experimental settings with three independent datasets with varying modalities. The precision, sensitivity, and specificity values for each independent dataset are shown in Table~\ref{tab2}, \ref{tab3}, \ref{tab4}. We evaluated the models’ performances under three evaluation scenarios: (i) with pretraining and random initialization; (ii) with pretrained DCNN architectures, and (iii) with DCNN architectural family. In the analysis 95\% confidence interval (CI) value quantifies the observed differences in performance for significant assessment of reliability.

\subsection{Evaluation with Pretraining and Random Initialization}\label{subsec4}
\begin{figure}[ht]
  \centering
  \begin{tabular}{ c @{\hspace{1pt}} c @{\hspace{1pt}} c}
      \includegraphics[width=.49\columnwidth]{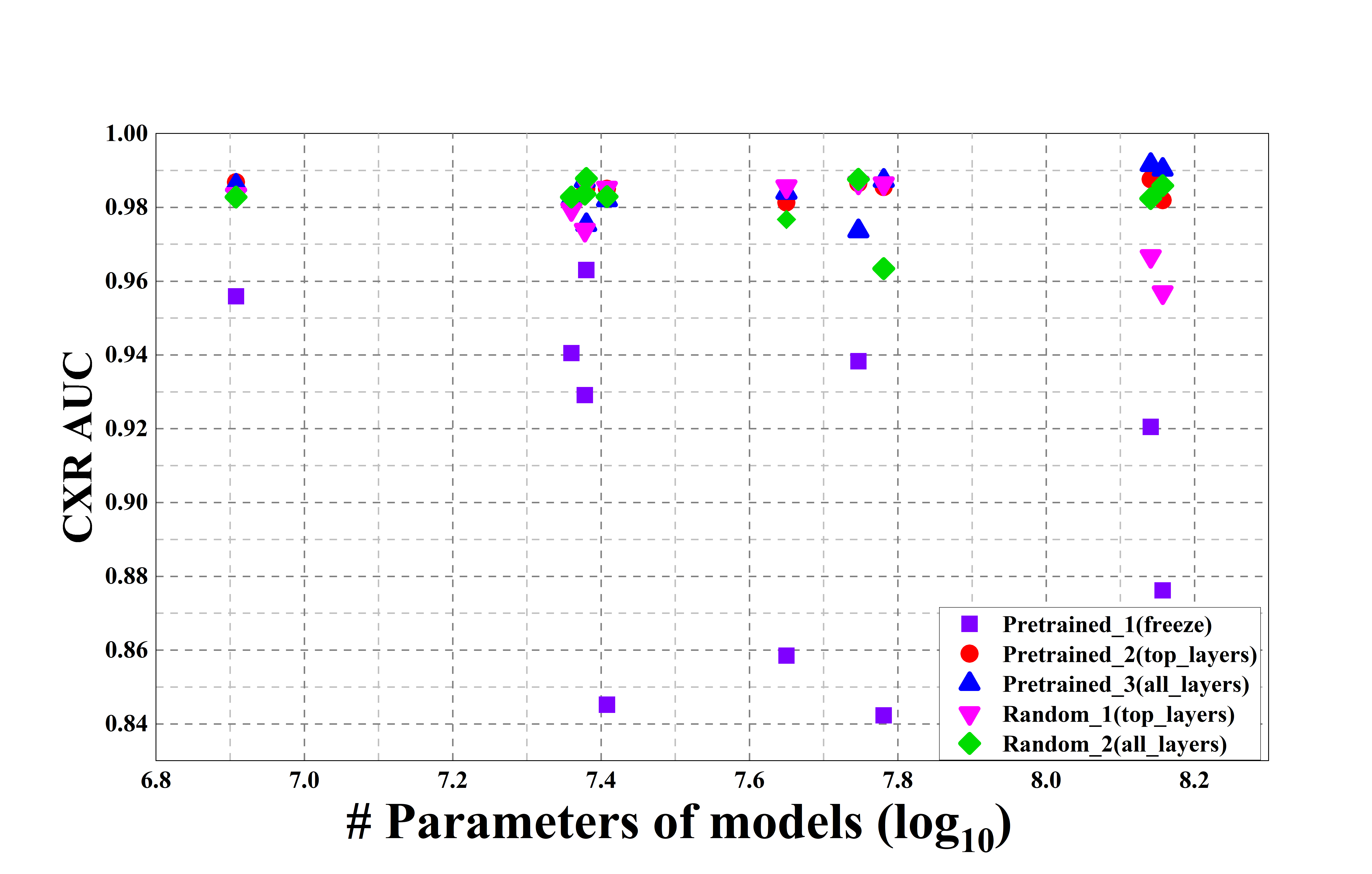} &
      \includegraphics[width=.49\columnwidth]{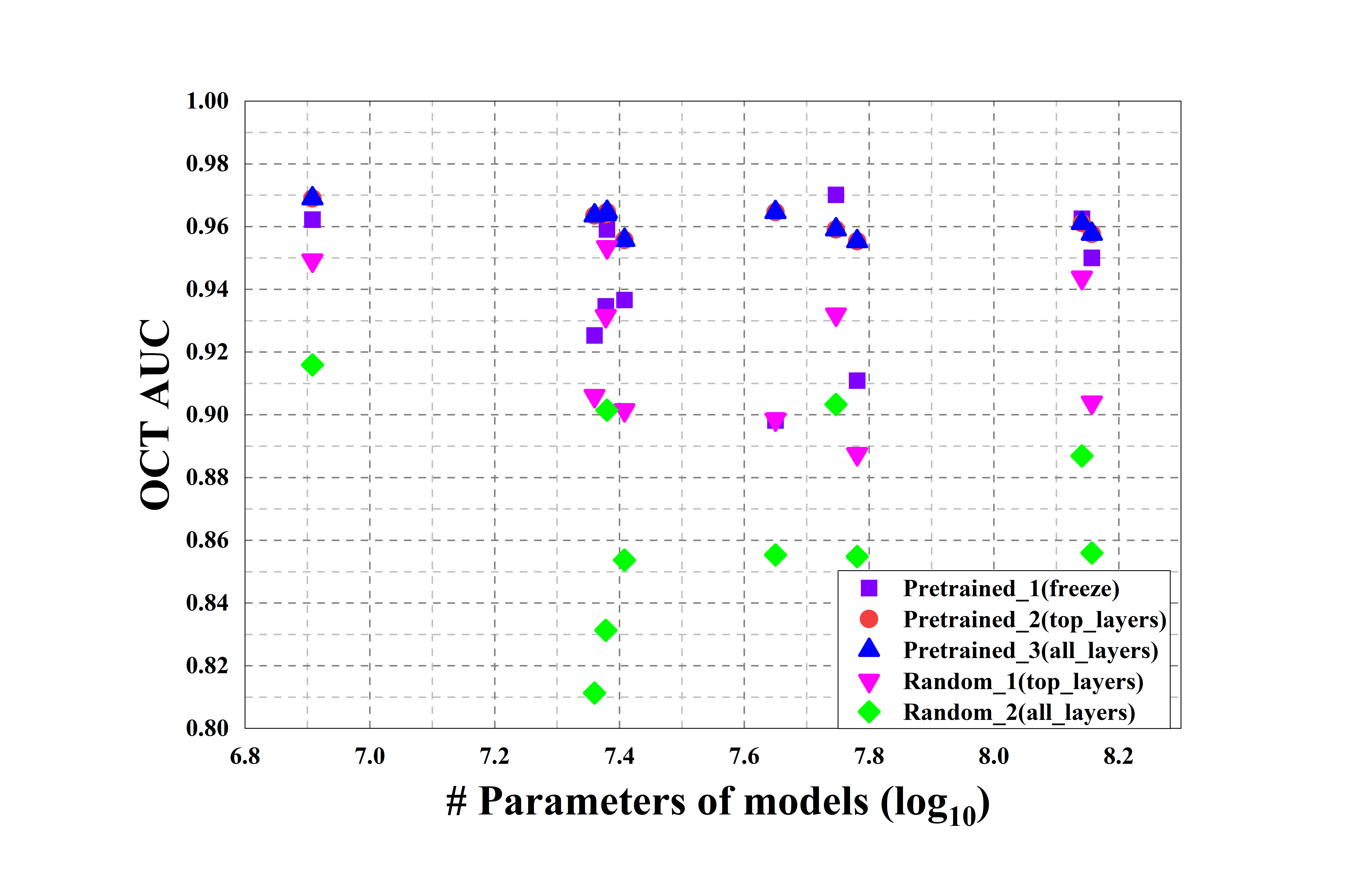} & \\
    \small (a)  &
    \small (b)  \\
        \includegraphics[width=.49\columnwidth]{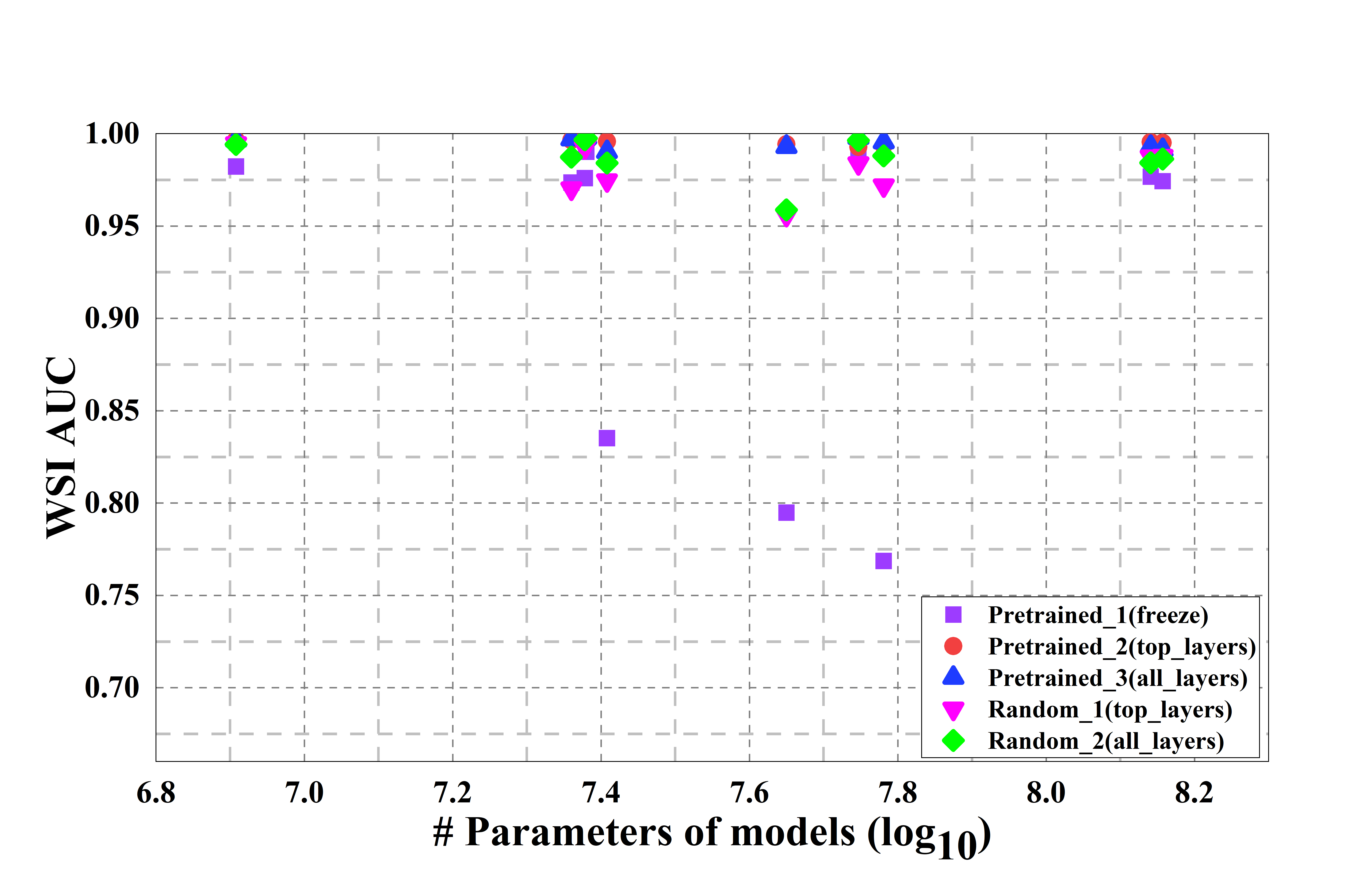} & \\
    \small (c) \\
    \end{tabular}
   \caption{Visual summary in scatter plots showing AUCs of (a) CXR (b) OCT, and (c) WSI of all DCNN pretrained models with pretraining and random initialization. The five different settings with the number of parameters represent the complexity of the models (VGG (16, 19), ResNet (50,
   101, 152),  Inceptionv3, Inception-ResNet-v2, Xception, and DenseNet (121, 201)). The values are represented in a logarithm scale (Table~\ref{tab1}).}\label{fig3}   
\end{figure}

From our observation comparing the impact of pretraining and random initialization we found pretraining helps increase the performance of the models while for some models, random initializations performed better. For CXR Figure~\ref{fig3} (a), VGG16, performance has been boosted with pretraining in training all layers of the network with an AUC 0.9916 followed by DenseNet121 with an AUC 0.9860. In the case of random initialization VGG16 and DenseNet121 have achieved AUC 0.9824, and 0.9828 respectively. A difference of only 0.0032 is observed under pretraining and random initialization.  For VGG16 a small difference in AUC of 0.0209 (95\% CI: 0.0139, 0.0279) was observed with pretraining and random initialization in training only the top layers of the networks. For VGG19 a difference of 0.0250 (95\% CI: 0.0061, 0.0374) was observed. For ResNet models, pretraining and random initialization showed small differences in performances. In pretraining ResNet152 gained the highest performance, while in random initialization ResNet50 gained performance. For ResNet152, a difference of 0.239 (95\% CI: 0.0129, 0.0608) was seen between pretraining and random initialization in training all layers of the networks. For the rest of the models, the performance in AUC with pretraining and random initializations is comparable under different settings.

In the case of OCT multiclass classification Figure~\ref{fig3} (b), it is observed pretraining helps increase the performance of DenseNet121, while random initialization helps DenseNet201. InceptionResNetV2 obtained the highest performance in training the network with only the top layers of the networks. In random initialization, VGG16 and DenseNet201 have a boost in performance with an AUC of 0.9437, and 0.9534 with DeneseNet201 obtaining the highest performance.  Compared to training all layers of the networks, in random initialization, training with only the top layers achieved lower performance. DenseNet121 achieved the highest performance with an AUC of 0.9113. With random initialization, in training all layers of the network and only the top layers a difference of 0.0421 is observed between the highest-performing model, while a difference of 0.0011 is observed with pretraining. In all the training settings, ResNet models showed inferior performance compared to other models. 

In histopathology WSI LC25000 multiclass classification Figure~\ref{fig3} (c), Inceptionv3 performed better over all other models with pretraining, followed by Xception, DenseNet121 and DenseNet201. While a difference of 0.0003 is observed between DenseNet121 and DenseNet201. Between the best and low-performing models, there is a difference of 0.0052 (95\% CI: 0.0034, 0.0157). With random initialization, DenseNet121 showed the highest performance with an AUC of 0.9948 (95\% CI: 0.9820, 1.0075), followed by DenseNet201 and Inceptionv3. ResNet101 shows the lowest performance. However, on comparing the performance with individual architecture the same is not followed as in the case of CXR and OCT.

\subsection{Evaluation with Pretrained DCNN architectures}\label{subsec5}
With pretraining, we observed that the largest and smallest models show slight differences in performance. For CXR DenseNet121, the smallest model has outperformed its larger peers. VGG16 with reduced parameters than VGG19 has better performance. When pretraining only the top layers of the networks, InceptionResNetv2 has performed better with an AUC of 0.9865 (95\% CI: 0.9688, 1.0044), (95\%CI: 0.9663, 1.0069) and 0.9876 (95\% CI: 0.9654, 1.0099). On the other hand, with random initialization VGG19 and ResNet152 fell short with an AUC of 0.9569 (95\% CI: 0.9621, 0.9646) and 0.9634 (95\% CI: 0.9621, 0.9646), respectively. Among the largest and the smallest architectures, there is an increase of 0.0049 (95\% CI: 0.0048, 0.0353), 0.0043 (95\% CI: 0.0015, 0.0240) with pretraining, while with random initializations, there is an increase of 0.0266 and 0.0031. 

From Figure~\ref{fig3} (b), in the case of OCT, a similar effect is observed but there exists variation in performance with the architecture of the models compared to CXR classification. With pretraining InceptionResNetV2 we achieved the highest performance with an AUC of 0.9701 (95\% CI: 0.9425, 0.9978) while ResNet101 showed the least performance at 0.8982 (95\% CI: 0.8796, 0.9168) in training only top layers of the networks. In training all layers of the networks, DenseNet121 showed the highest performance with an AUC of 0.9690 (95\% CI: 0.9537, 0.9844), and ResNet152 with larger parameters achieved a lower performance of 0.9553 (95\% CI: 0.9444, 0.9662). With random initialization, DenseNet201 achieved the highest performance with an AUC of 0.9534 (95\% CI: 0.9407, 0.9660) in training all layers of the networks while ResNet152 achieved the least performance at 0.8875 (95\% CI: 0.8619, 0.9130). On the other hand, DenseNet121 has the highest performance with 0.9159 (95\% CI: 0.9044, 0.9275) and Xception has the least performance 0.8133 (95\% CI: 0.7977, 0.8249). Among the highest and least performing models a performance difference of 0.1046.

In histopathological LC25000 from Figure~\ref{fig3} (c), Inceptionv3 achieved the highest performance compared to other architectures with larger parameters with an AUC of 0.9980 (95\% CI: 0.9954, 1.0006) followed by DenseNet201 with an AUC of 0.9968 (95\% CI: 0.9904, 1.0033) in training all layers of the networks. Training only the top layers of the network InceptionResNetV2 and DenseNet201 achieved similar performance with an AUC of 0.9979. VGG19 showed the lowest performance. Whereas for random initialization, training all the layers of the networks, DenseNet121 achieved the highest performance with an AUC of 0.9948 (95\% CI: 0.9820, 1.0075) and ResNet101 the least performance at 0.9559 (95\% CI: 0.9075, 1.0043). In training only the top layers of the networks, Inceptionv3 achieved the highest performance with an AUC of 0.9968 (95\% CI: 0.9966, 0.9969) and ResNet101 with the least performance.

\begin{sidewaystable}
\caption{Precision, Recall, Specificity of all ten models under study both with and without pretraining for CXR binary image classification. The table below shows the values of their respective scores with 95\% confidence intervals, CI.}\label{tab2}
\begin{tabular*}{\textheight}{@{\extracolsep\fill}lcccccc}
\toprule%
& \multicolumn{3}{@{}c@{}}{Transfer (Fine-tune)}& \multicolumn{2}{@{}c@{}}{Random Initialization} \\\cmidrule{2-4}\cmidrule{5-6}%
\begin{tabular}[c]{@{}l@{}}Model \\ Architecture\end{tabular} & \multicolumn{1}{c}{\begin{tabular}[c]{@{}c@{}}Pretrained\_1\\ (freeze)\\ (95\%  CI)\end{tabular}} & \multicolumn{1}{c}{\begin{tabular}[c]{@{}c@{}}Pretrained\_2\\ (top\_layers)\\ (95\% CI)\end{tabular}} & \multicolumn{1}{c}{\begin{tabular}[c]{@{}c@{}}Pretrained\_3\\ (all\_layers)\\ (95\% CI)\end{tabular}} & \multicolumn{1}{c}{\begin{tabular}[c]{@{}c@{}}Random\_1\\ (top\_layers)\\ (95\% CI)\end{tabular}} & \multicolumn{1}{c}{\begin{tabular}[c]{@{}c@{}}Random\_2\\ (all\_layers)\\ (95\% CI)\end{tabular}} \\ 
\cmidrule{2-6}%
\multicolumn{6}{c}{Precision (95\% CI)} \\
\midrule
VGG16 & 0.8862 (0.8488, .9236) & 0.9287 (0.8989, 0.9586) & 0.9278(0.9024, 0.9532) & 0.8789(0.8662, 0.8916) & 0.9218 (0.8957, 0.9478) \\
VGG19 & 0.8159 (0.7740, .8579) & 0.9268(0.9090, 0.9446)  & 0.9277(0.7346, 1.1208)  & 0.8716(0.8366, 0.9065) & 0.9302 (0.8895, 0.9709) \\
ResNet50 & 0.6745 (0.6529, .6961) & 0.9296 (0.9175, 0.9417)  & 0.8980(0.7030, 1.0930)  & 0.9091 (0.8685, 0.9498) & 0.9291 (0.8516, 1.0066) \\
ResNet101 & 0.7041 (0.7003, .7079) & 0.8972 (0.8165, 0.9778)  & 0.8733(0.7983, 0.9483)  & 0.9064 (0.7984, 1.0144) & 0.9144 (0.8839, 0.9449) \\
ResNet152 & 0.6568 (.5990, .7146) & 0.8946 (0.8717, 0.9175)  & 0.9075(0.8433, 0.9716)  & 0.9103 (0.7337,1.0144) & 0.9035(0.8914, 0.9155) \\
Inceptionv3 & 0.8321 (0.7730, .8912) & 0.9090 (0.8830, 0.9351)  & 0.8947(0.7587, 1.0306)  & 0.8985(0.7740, 1.0230) & 0.9061 (0.8419, 0.9703) \\
Inception-ResNet-v2\ & 0.8548 (0.8453, .8643) & 0.9093 (0.8953, 0.9232)  & 0.8921(0.8223, 0.9620)  & 0.9317 (0.9305, 0.9330) & 0.9312 (0.7927, 1.0697) \\
Xception & 0.8734 (0.8626, 0.8842) & 0.8827 (0.8096, 0.9558)  & 0.8772 (0.7546, 0.9998)  & 0.9047 (0.8602, 0.9491) & 0.9305 (0.8555, 1.0054) \\
DenseNet121 & 0.8782 (0.8675, 0.8891) & 0.9104 (0.8418,0.9790)  & 0.8750 (0.7566, 0.9945)  & 0.9349 (0.8600, 1.0099) & 0.9441 (0.9428, 0.9453) \\
DenseNet201 & 0.8522 (0.8033, .9011) & 0.9167 (0.9027, 0.9306)  & 0.8578 (0.7066, 1.0090)  & 0.9254(0.9101, 0.9406) & 0.9446 (0.9440, 0.9453) \\
\midrule
\multicolumn{6}{c}{Sensitivity (95\% CI)} \\
\midrule
VGG16 & 0.8952 (0.8908, 0.8996) & 0.9794 (0.9629, 0.9960) & 0.9880 (0.9391, 1.0369)	& 0.9794 (0.9794, 0.9794) & 0.9871 (0.9541, 1.0202) \\
VGG19 & 0.9184 (0.8206, 1.0163) & 0.9743(0.9254, 1.0230) & 0.9837 (0.9355, 1.0320) & 0.9743 (0.9419, 1.0067) & 0.9914 (0.9590, 1.0238) \\
ResNet50 & 0.9557 (0.9443, 0.9672) & 0.9709 (0.9385, 1.0033) & 0.9914 (0.9590, 1.0238) & 0.9812 (0.9494, 1.0129) & 0.9735 (0.9080, 1.0389) \\
ResNet101 & 0.9376 (0.9198, 0.9554) & 0.9837 (0.9672, 1.0003) & 0.9957 (0.9957, 0.9957)	& 0.9846 (0.9680, 1.0011) &	0.9760 (0.9430, 1.0090) \\
ResNet152 & 0.9583 (0.8491, 1.0676) & 0.9837(0.9539, 1.0187) & 0.9888 (0.9399, 1.0378) & 0.9880 (0.9721, 1.0039) & 0.9598 (0.9439, 0.9757) \\
Inceptionv3 & 0.9532 (0.8719, 1.0345) & 0.9820 (0.9662, 0.9979) & 0.9946 (0.9330, 1.0563) & 0.9820 (0.9007, 1.0634) & 0.9811 (0.9487,1.01335) \\
Inception-ResNet-v2\ & 0.9465 (0.9299, 0.9630) & 0.9898 (0.9898, 0.9898) & 0.9794 (0.9636, 0.9953) & 0.9803 (0.9638, 0.9968) & 0.9820 (0.9172, 1.0468) \\
Xception & 0.9236 (0.9236, 0.9236) & 0.9376 (0.92821.2470) & 0.9948 (0.9624, 1.0272) & 0.9581 (0.9257, 0.9905) & 0.9829 (0.9498, 1.0159) \\
DenseNet121 & 0.9549 (0.9384, 0.9715) & 0.9871 (0.9713, 1.0330) & 0.9947 (0.9922, 0.9972) & 0.9820 (0.9503, 1.0138) & 0.9820 (0.8168, 1.1472) \\
DenseNet201 & 0.9646 (0.9643, 0.9643) & 0.9872 (0.9872, 0.9872) & 0.9931 (0.9918, 0.9943) & 0.9846 (0.9681, 1.0011) & 0.9786 (0.9621, 0.9951) \\
\midrule
\multicolumn{6}{c}{Specificity (95\% CI)} \\
\midrule
VGG16 & 0.8862 (0.8488, 0.9237) & 0.8946 (0.8400, 2.6480) & 0.8949 (0.8339, 0.9559) & 0.7749 (0.7482, 0.8019)	& 0.8830 (0.8284, 0.9376) \\
VGG19 & 0.8159 (0.7740, 0.8579) & 0.8922 (0.8648, 2.6408) &	0.8867 (0.8998, 1.2736) & 0.7407 (0.4414, 1.0399)	& 0.8947 (0.8133, 0.9761) \\
ResNet50 & 0.6745 (0.6529, 0.661) & 0.8924 (0.8651, 2.6415) & 0.8435 (0.5061, 1.1808) & 0.8348 (0.7534, 0.9162)	& 0.9004 (0.8026, 0.9982) \\
ResNet101 & 0.7041 (0.7003, 0.7079) & 0.8389 (0.6756, 1.0022) & 0.7961 (0.6328, 0.9593) & 0.8362 (0.5103, 1.1621)	& 0.8653 (0.8107, 0.9200) \\
ResNet152 & 0.6568 (0.5990, 0.7146) & 0.8332 (0.7786, 0.8878) & 0.8581 (0.7221, 0.9941) & 0.8376 (0.4842, 1.1909)	& 0.8561 (0.8288, 0.8834) \\
Inceptionv3 & 0.8321 (0.7730, 0.8912) & 0.8638 (0.8092, 0.9184)	& 0.8387 (0.5668, 1.1106) & 0.8148 (0.5435, 1.0860) & 0.8519(0.7159, 0.9878) \\
Inception-ResNet-v2\ & 0.8548 (0.8453, 0.8643) & 0.8556 (0.8289, 0.8823) & 0.8441 (0.7081, 0.9800)= & 0.8803 (0.8803, 0.8803) & 0.8941 (0.6222, 1.1660) \\
Xception & 0.8734 (0.8639, 0.8829) & 0.8246 (0.7433, 0.9059) & 0.8047 (0.5334, 1.0760) & 0.8319 (0.7506, 0.9133) & 0.8981 (0.7621, 1.0340) \\
DenseNet121 & 0.8783 (0.8675, 0.8891) & 0.8684 (0.7330, 1.0037)	& 0.7899 (0.5187, 1.0612) & 0.8860 (0.7507, 1.0213) & 0.9151 (0.9151, 0.9151) \\
DenseNet201 & 0.8522 (0.8033, 0.9011) & 0.8733 (0.8460, 0.9006) & 0.7730 (0.4387, 1.0773) & 0.8675 (0.8402, 0.8948)	& 0.9171 (0.9171, 0.9171) \\
\bottomrule
\end{tabular*}
\end{sidewaystable}

\begin{sidewaystable}
\caption{Precision, Recall, Specificity of all ten models under study both with and without pretraining for OCT multiclass image classification. The table below shows the values of their respective scores with 95\% confidence intervals, CI.}\label{tab3}
\begin{tabular*}{\textheight}{@{\extracolsep\fill}lcccccc}
\toprule{1-6}%
& \multicolumn{3}{@{}c@{}}{Transfer (Fine-tune)}& \multicolumn{2}{@{}c@{}}{Random Initialization} \\\cmidrule{2-4}\cmidrule{5-6}%
\begin{tabular}[c]{@{}l@{}}Model \\ Architecture\end{tabular} & \multicolumn{1}{c}{\begin{tabular}[c]{@{}c@{}}Pretrained\_1\\ (freeze)\\ (95\%  CI)\end{tabular}} & \multicolumn{1}{c}{\begin{tabular}[c]{@{}c@{}}Pretrained\_2\\ (top\_layers)\\ (95\% CI)\end{tabular}} & \multicolumn{1}{c}{\begin{tabular}[c]{@{}c@{}}Pretrained\_3\\ (all\_layers)\\ (95\% CI)\end{tabular}} & \multicolumn{1}{c}{\begin{tabular}[c]{@{}c@{}}Random\_1\\ (top\_layers)\\ (95\% CI)\end{tabular}} & \multicolumn{1}{c}{\begin{tabular}[c]{@{}c@{}}Random\_2\\ (all\_layers)\\ (95\% CI)\end{tabular}} \\ 
\cmidrule{2-6}%
\multicolumn{6}{c}{Precision (95\% CI)} \\
\midrule
VGG16 & 0.7343 (0.6830, 0.7856) & 0.9433 (0.9397, 0.9469) & 0.9491 (0.9397, 0.9587)	& 0.8558 (0.8367, 0.8748)	& 0.9275 (0.9051, 0.9499) \\
VGG19 & 0.7227 (0.7151, 0.7302)	& 0.9383 (0.9204, 0.9562) & 0.9442 (0.9347, 0.9536)	& 0.8017 (0.7860, 0.8173)	& 0.8900 (0.8088, 0.9712)\\
ResNet50 & 0.4690 (0.4624, 0.4756) & 0.9258 (0.9163, 0.9353) & 0.9425 (0.9636, 0.9487) & 0.8400 (0.7915, 0.8885)	& 0.8850 (0.8822, 0.8894) \\
ResNet101 & 0.4490 (0.4192, 0.4788) & 0.8800 (0.8585, 0.9015) & 0.9533 (0.9438, 0.9628) & 0.8150 (0.7715, 0.8585) & 0.8592 (0.8004, 0.9179) \\
ResNet152 & 0.4470 (0.4371, 0.4569) & 0.8900 (0.8776, 0.9024) & 0.9442 (0.9285, 0.9598)	& 0.8075 (0.7403, 0.8747)	& 0.8625 (0.8302, 0.8948) \\
Inceptionv3 & 0.8542 (0.8447, 0.8637) & 0.9192 (0.9097, 0.9287)	& 0.9542 (0.9447, 0.9637) & 0.8050 (0.7835, 0.8265) & 0.9133 (0.8977, 0.9289) \\
Inception-ResNet-v2\ & 0.8275 (0.8111, 0.8439) & 0.9533 (0.9497, 0.9569) & 0.9453 (0.9165, 0.9752) & 0.8842 (0.8769, 0.8916) & 0.9175 (0.9051, 0.9299) \\
Xception & 0.8167 (0.7977, 0.8356) & 0.9092 (0.9056, 0.9127) & 0..9533 (0.9462, 0.9605) & 0.7767 (0.7486, 0.8047)	& 0.8908 (0.8813, 0.9003) \\
DenseNet121 & 0.8627 (0.8564, 0.8689) & 0.9533 (0.9404, 0.9663)	& 0.9600 (0.9376, 0.9824) & 0.9000 (0.8876, 0.9124) & 0.9375 (0.9104, 0.9646) \\
DenseNet201 & 0.8558 (0.8402, 0.8715) & 0.9458 (0.9259, 0.9658)	& 0.9533 (0.9462, 0.9605) & 0.8025 (0.8701, 0.8949) & 0.9417 (0.9217, 0.9616) \\
\midrule
\multicolumn{6}{c}{Sensitivity (95\% CI)} \\
\midrule
VGG16 & 0.7343 (0.6830, 0.7856)	& 0.9417 (0.9322, 0.9512) &	0.9343 (0.9303, 0.9382)	& 0.9150 (0.8865, 0.9435)	& 0.8308 (0.8272, 0.8344) \\
VGG19 & 0.7227 (0.7151, 0.7302) & 0.9367 (0.9272, 0.9452) & 0.9258 (0.8971, 0.9545) & 0.8567 (0.7284, 0.9849)	& 0.7842 (0.7747, 0.7937) \\
ResNet50 & 0.4690 (0.4624, 0.4756) & 0.9350 (0.9242, 0.9458) & 0.9042 (0.8807, 0.9277) & 0.8517 (0.8422, 0.8612)	& 0.7808 (0.7231, 0.8385) \\
ResNet101 & 0.4490 (0.4192, 0.4788) & 0.9475 (0.9416, 0.9537) & 0.8475 (0.8146, 0.8804) & 0.8280 (0.7596, 0.8821) & 0.7808 (0.7489, 0.8127) \\
ResNet152 & 0.4470 (0.4371, 0.4569) & 0.9342 (0.9185, 0.9498) & 0.8675 (0.8511, 0.8839)	& 0.8292 (0.7892, 0.8691)	& 0.7833 (0.7454, 0.8213) \\
Inceptionv3 & 0.8200 (0.7929, 0.8471) & 0.9450 (0.9326, 0.9574)	& 0.9008 (0.8913, 0.9103) & 0.8958 (0.8740, 0.9176) & 0.7542 (0.7084, 0.7999) \\
Inception-ResNet-v2\ & 0.7758 (0.7507, 0.8009) & 0.9392 (0.9073, 0.9710) & 0.9483 (0.9412, 0.9555) & 0.8992 (0.8802, 0.9181) & 0.8550 (0.8488, 0.8612) \\
Xception & 0.7728 (0.7559, 0.7898) & 0.9458 (0.9329, 0.9588) & 0.8900 (0.8792, 0.9007) & 0.8600 (0.8414, 0.8786) & 0.7158 (0.6940, 0.7376) \\
DenseNet121 & 0.8167 (0.7873, 0.8460) & 0.9558 (0.9299, 0.9817)	& 0.9525 (0.9137, 0.9913) & 0.9292 (0.8927, 0.9656)	& 0.8742 (0.8552, 0.8931)\\
DenseNet201 & 0.8113 (0.7907, 0.8319) & 0.9475 (0.9551, 0.9599) & 0.9392 (0.9098, 0.9685) & 0.9300 (0.9136, 0.9464)	& 0.8500 (0.8252, 0.8748) \\
\midrule
\multicolumn{6}{c}{Specificity (95\% CI)} \\
\midrule
VGG16 & 0.9114 (0.8944, 0.9285) & 0.9776 (0.9762, 0.9791) & 0.9805 (0.9763, 0.9848)	& 0.9435 (0.9429, 0.9439)	& 0.9718 (0.9627, 0.9809) \\
VGG19 & 0.9076 (0.9051, 0.9101) & 0.9749 (0.9669, 0.9829) & 0.9784 (0.9759, 0.9809)	& 0.9279 (0.9259, 0.9301) & 0.9519 (0.9086, 0.9952) \\
ResNet50 & 0.8233 (0.8211, 0.8256) & 0.9683 (0.9605, 0.9761) & 0.9778 (0.9744, 0.9812) & 0.9268 (0.9072, 0.9465)	& 0.9507 (0.9476, 0.9538) \\
ResNet101 & 0.8163 (0.8063, 0.8263) & 0.9482 (0.9361, 0.9603) & 0.9823 (0.9798, 0.9848)	& 0.9276 (0.9163, 0.9389) & 0.9221 (0.8539, 0.9902) \\
ResNet152 & 0.8166 (0.8145, 0.8187)	& 0.9554 (0.9508, 0.9600) & 0.9776 (0.9722, 0.9829) & 0.9274 (0.9147, 0.9399)	& 0.9438 (0.9308, 0.9567) \\
Inceptionv3 & 0.9419 (0.9378, 0.9574) & 0.9673 (0.9638, 0.9708) & 0.9819 (0.9786, 0.9853) & 0.9157 (0.9111, 0.9202) & 0.9657 (0.9591, 0.9724) \\
Inception-ResNet-v2\ & 0.9269 (0.9176, 0.9361) & 0.9712 (0.9256, 1.0168) & 0.9796 (0.9689, 0.9902) & 0.9516 (0.9491, 0.9541) & 0.9659 (0.9599, 0.9719) \\
Xception & 0.9243 (0.9189, 0.9296) & 0.9626 (0.9598, 0.9655) & 0.9817 (0.9778, 0.9856) & 0.9053 (0.8980, 0.9126) & 0.9536 (0.9465, 0.9608) \\
DenseNet121 & 0.9424 (0.9392, 0.9456) & 0.9815 (0.9774, 0.9856) & 0.9845 (0.9769, 0.9921) & 0.9579 (0.9521, 0.9638) & 0.9758 (0.9631, 0.9885)\\
DenseNet201 & 0.9383 (0.9316, 0.9451) & 0.9795 (0.9721, 0.9867) & 0.9823 (0.9783, 0.9863) & 0.9499 (0.9413, 0.9587)	& 0.9768 (0.9705, 0.9831)\\
\bottomrule
\end{tabular*}

\end{sidewaystable}

\begin{sidewaystable}
\caption{Precision, Recall, Specificity of all ten models under study both with and without pretraining for histopathology microscopy WSI multiclass image classification. The table below shows the values of their respective scores with 95\% confidence intervals, CI.}\label{tab4}
\begin{tabular*}{\textheight}{@{\extracolsep\fill}lcccccc}
\toprule{1-6}%
& \multicolumn{3}{@{}c@{}}{Transfer (Fine-tune)}& \multicolumn{2}{@{}c@{}}{Random Initialization} \\\cmidrule{2-4}\cmidrule{5-6}%
\begin{tabular}[c]{@{}l@{}}Model \\ Architecture\end{tabular} & \multicolumn{1}{c}{\begin{tabular}[c]{@{}c@{}}Pretrained\_1\\ (freeze)\\ (95\%  CI)\end{tabular}} & \multicolumn{1}{c}{\begin{tabular}[c]{@{}c@{}}Pretrained\_2\\ (top\_layers)\\ (95\% CI)\end{tabular}} & \multicolumn{1}{c}{\begin{tabular}[c]{@{}c@{}}Pretrained\_3\\ (all\_layers)\\ (95\% CI)\end{tabular}} & \multicolumn{1}{c}{\begin{tabular}[c]{@{}c@{}}Random\_1\\ (top\_layers)\\ (95\% CI)\end{tabular}} & \multicolumn{1}{c}{\begin{tabular}[c]{@{}c@{}}Random\_2\\ (all\_layers)\\ (95\% CI)\end{tabular}} \\ 
\cmidrule{2-6}%
\multicolumn{6}{c}{Precision (95\% CI)} \\
\midrule
VGG16 & 0.9630 (0.9492, 0.9768)	& 0.9944 (0.9849, 1.0038) & 0.9767 (0.9193, 1.0340)	& 0.9822 (0.9649, 0.9995) & 0.9800 (0.9716, 0.9883) \\
VGG19 & 0.9655 (0.9608, 0.9702) & 0.9911 (0.9864, 0.9958) & 0.9877 (0.9751, 1.0004) & 0.9833 (0.9751, 0.9915) & 0.9833 (0.9689, 0.9976)\\
ResNet50 & 0.7733 (0.7589, 0.7877) & 0.9955 (0.9908, 1.0002) & 0.9866 (0.9579, 1.0153) & 0.9655 (0.9389, 0.9922) & 0.9788 (0.9522, 1.0055) \\
ResNet101 & 0.7222 (0.7175, 0.7269) & 0.9911 (0.9815, 1.0006) & 0.9922 (0.9827, 1.0017)	& 0.9511 (0.9176, 0.9846) & 0.9466 (0.8098, 1.0834) \\
ResNet152 & 0.6889 (0.6842, 0.6936) & 0.9922 (0.9875, 0.9969) & 0.9955 (0.9829, 1.0083)	& 0.9600 (0.9434, 0.9765)	& 0.9844 (0.9797, 0.9891) \\
Inceptionv3 & 0.9666 (0.9583, 0.9749) & 0.9989 (0.9939, 1.0003)	& 0.9966 (0.9883, 1.0049) & 0.9900 (0.9816, 0.9983)	& 0.9953 (0.9894, 1.0012) \\
Inception-ResNet-v2 & 0.9844 (0.9749, 0.9939) & 0.9900 (0.9681, 1.0118) & 0.9983 (0.9938, 1.0027) & 0.9734 (0.9588, 0.9879)	& 0.9955 (0.9908, 1.0002) \\
Xception & 0.9623 (0.9526, 0.9719) & 0.9998 (0.9991, 1.0006) & 0.9978 (0.9882, 1.0074) & 0.9578 (0.9451, 0.9704) & 0.9839 (0.9876, 0.9869) \\
DenseNet121 & 0.9755 (0.9708, 0.9802) & 0.9966 (0.9893, 1.0049)	& 0.9987 (0.9942, 1.0032) & 0.9944 (0.9771, 1.0117)	& 0.9933 (0.9714, 1.0152) \\
DenseNet201 & 0.9855 (0.9807, 0.9902) & 0.9978 (0.9882, 1.0074)	& 0.9977 (0.9929, 1.0026) & 0.9933 (0.9851, 1.0014) & 0.9978 (0.9882, 1.0074) \\
\midrule
\multicolumn{6}{c}{Sensitivity (95\% CI)} \\
\midrule
VGG16 & 0.9703 (0.9696, 0.9710) & 0.9844 (0.9452, 1.0235) & 0.9933 (0.9789, 1.0077)	& 0.9855 (0.9759, 0.9951) & 0.9789 (0.9739, 0.9837) \\
VGG19 & 0.9660 (0.9629, 0.9690) & 0.9922 (0.9875, 0.9969) & 0.9889 (0.9762, 1.0015)	& 0.9844 (0.9797, 0.9891) & 0.9811 (0.9684, 0.9937) \\
ResNet50 & 0.7800 (0.7716, 0.7883) & 0.9955 (0.9908, 1.0002) & 0.9877 (0.9638, 1.0117) & 0.9666 (0.9447, 0.9886) & 0.9789 (0.9598, 0.9979) \\
ResNet101 & 0.7266 (0.7183, 0.7349)	& 0.9933 (0.9851, 1.0015) & 0.9922 (0.9827, 1.0016)	& 0.9411 (0.8742, 1.0079) & 0.9455 (0.7970, 1.0939) \\
ResNet152 & 0.6933 (0.6850, 0.7017)	& 0.9922 (0.9875, 0.9969) & 0.9966 (0.9883, 1.0049)	& 0.9622 (0.9413, 0.9831)	& 0.9844 (0.9797, 0.9891) \\
Inceptionv3 & 0.9700 (0.9616, 0.9783) & 0.9989 (0.9939, 1.0037)	& 0.9978 (0.9882, 1.0074) & 0.9911 (0.9864, 0.9958)	& 0.9989 (0.9939, 1.0037) \\
Inception-ResNet-v2\ & 0.9866 (0.9783, 0.9949) & 0.9922 (0.9731, 1.0112) & 0.9988 (0.9941, 1.0035) & 0.9812 (0.9761, 09863) & 0.9955 (0.9908, 1.0002) \\
Xception & 0.9654 (0.9532, 0.9776) & 0.9999 (0.9993, 1.0004) & 0.9978 (0.9882, 1.0074) & 0.9677 (0.9471, 0.9883) & 0.9833 (0.9751, 0.9915) \\
DenseNet121 & 0.9755 (0.9708, 0.9802) & 0.9966 (0.9883, 1.0049) & 0.9995 (0.9982, 1.0008) & 0.9944 (0.9771, 1.0117)	& 0.9933 (0.9714, 1.0152)\\
DenseNet201 & 0.9877 (0.9829, 0.9926) & 0.9978 (0.9882, 1.0078)	& 0.9977 (0.9929, 1.0026) & 0.9947 (0.9867, 1.0026)	& 0.9978 (0.9882, 1.0074) \\
\midrule
\multicolumn{6}{c}{Specificity (95\% CI)} \\
\midrule
VGG16 & 0.9830 (0.9799, 0.9860) & 0.9970 (0.9931, 1.0008) & 0.9954 (0.9926, 0.9982) & 0.9917 (0.9859, 0.9975) & 0.9910 (0.9840, 0.9981) \\
VGG19 & 0.9828 (0.9818, 1.1645)	& 0.9956 (0.9947, 0.9969) & 0.9942 (0.9899, 0.9986) & 0.9921 (0.9902, 0.9939) & 0.9921 (0.9885, 0.9956) \\
ResNet50 & 0.8895 (0.8863, 0.8927) & 0.9970 (0.9903, 1.0037) & 0.9935 (0.9819, 1.0051) & 0.9833 (0.9713, 0.9952)	& 0.9880 (0.9768, 0.9992) \\
ResNet101 & 0.8618 (0.8603, 0.8633) & 0.9962 (0.9929, 0.9994) & 0.9954 (0.9914, 0.9994)	& 0.9708 (0.9374, 1.0042)	& 0.9705 (0.9039, 1.0371) \\
ResNet152 & 0.8448 (0.8439, 0.8457)	& 0.9957 (0.9932, 0.9982) & 0.9970 (0.9929, 1.0011) & 0.9815 (0.9729, 0.9901)	& 0.9870 (0.9676, 1.0065) \\
Inceptionv3 & 0.9841 (0.9818, 0.9864) & 0.9980 (0.9954, 1.0006) & 0.9959 (0.9888, 1.0029) & 0.9950 (0.9916, 0.9985)	& 0.9974 (0.9957, 0.9991)\\
Inception-ResNet-v2\ & 0.9927 (0.9894, 0.9960) & 0.9952 (0.9866, 1.0036) & 0.9962 (0.9831, 1.0092) & 0.9894 (0.9884, 0.9903) & 0.9952 (0.9823, 1.0080) \\
Xception & 0.9851 (0.9748, 0.9954) & 0.9991 (0.9971, 1.0011) & 0.9959 (0.9877, 1.0041) & 0.9802 (0.9712, 0.9819) & 0.9864 (0.9651, 1.0077) \\
DenseNet121 & 0.9883 (0.9866, 0.9901) & 0.9977 (0.9952, 1.0002)	& 0.9988 (0.9979, .9996) & 0.9965 (0.9879, 1.0052)	& 0.9945 (0.9844, 1.0046)\\
DenseNet201 & 0.9935 (0.9926, 0.9945) & 0.9979 (0.9935, 1.0023)	& 0.9986 (0.9968, 1.0003) & 0.9951 (0.9884, 1.0018) & 0.9984 (0.9959, 1.0009)\\
\bottomrule
\end{tabular*}
\end{sidewaystable}

\subsection{Evaluation with Pretrained DCNN architectural family}\label{subsec6}
Here, the architectural family represents a model family with an increased number of parameters such as VGG (VGG16, VGG19), ResNet (ResNet50, ResNet101, ResNet152), DenseNet (DenseNet121, DenseNet201). In the case of CXR Figure~\ref{fig3} (a), the VGG family, with pretraining VGG16 performed better with AUCs of 0.9876 (95\% CI: 0.9781, 0.9971) and 0.9916 (95\% CI: 0.9897, 0.9935) in training top layers and all layers of the network respectively. While with random initializations, VGG19 has performed better with an AUC of 0.9859 (95\% CI: 0.9688, 1.0031) in training all layers of the network. With ResNet families, ResNet152 showed superior performance with AUCs of 0.9855 (95\% CI: 0.9798, 0.9913) and 0.9873 (95\% CI: 0.9492, 1.0254) with pretraining. Meanwhile, ResNet50 performed better with an AUC of 0.9829 (95\% CI: 0.9594, 1.0064) under random initializations, in training all layers of the networks. Within the DenseNet family, DenseNet121 performed better with AUCs 0.9868 (95\% CI: 0.9804, 0.9932), and 0.9860 (95\% CI: 0.9663, 1.0057) with pretraining. DenseNet201, on the other hand, demonstrated improved performance with AUCs of 0.9865. 

For OCT Figure~\ref{fig3} (b), with pretraining ResNet family showed the least performance compared to other architectures. However, within the ResNet family, ResNet50 showed the highest performance with an AUC of 0.9366 (95\% CI: 0.9209, 0.9523) compared to two larger architectural counterparts ResNet101, and ResNet152. However, ResNet101 achieved the highest performance in training all network layers with an AUC of 0.9646 (95\% CI: 0.9595, 0.9697) under pretraining. Within the highest-performing model, VGG16 gained the highest performance. DenseNet121, and DensNet201 have a performance difference of 0.0032 (95\% CI: 0.0032, 0.0096) with pretraining, with DenseNet121 gaining the highest performance with an AUC 0.9622 (95\% CI: 0.9539, 0.9706) in training top layers of the networks. With random initialization, VGG16 attained the highest performance with an AUC of 0.9437 (95\% CI: 0.9252, 0.9622) compared to VGG19. Among the ResNet family, in training all layers of the network, ResNet50 obtained the highest performance with an AUC of 0.9015 (95\% CI: 0.8954, 0.9077) and ResNet152 attained the least performance with an AUC of 0.8875 (95\% CI: 0.8619, 0.9130). While DenseNet201 achieved the highest performance with an AUC of 0.9534 (95\% CI: 0.9407, 0.9660) compared to DenseNet121. 

In histopathological LC25000 image Figure~\ref{fig3} (c), with pretraining within the VGG family VGG16 achieved a performance boost with an AUC of 0.9955 (95\% CI: 0.9895, 1.0014) and 0.9930 (95\% CI: 0.9885, 0.9976) in training all layers and only top layers of the of the network respectively. In the case of ResNet family, ResNet152 achieved the highest performance with an AUC of 0.9955 (95\% CI: 0.9955, 0.9961) with pretraining in training only the top layers of the network, while ResNet50 achieved the highest performance with an AUC of 0.9958 (95\% CI: 0.9955, 0.9961) in training all layers of the network. ResNet101 showed the least performance. Among the DenseNet family, DenseNet201 achieved the highest performance. For random initialization, VGG19 achieved the highest performance with an AUC of 0.9879 (95\% CI: 0.9850, 0.9908). Among the ResNet family, ResNet50 and ResNet152 achieved the highest performance with an AUC of 0.9746 and 0.9880 in training all layers of the networks and only the top layers of the network respectively. ResNet101 shows the least performance. For the DenseNet family, DenseNet121 achieved the highest performance with an AUC of 0.9948 and 0.9959 compared to its larger counterpart. 

\section{Discussion}\label{sec5}

In this work, the use of different pretrained DCNNs on CXR, OCT, and histopathology WSI for binary and multiclass disease classification and their performance on these three specific datasets are carefully analyzed and reported. A general question that arises on using pretrained models with TL is whether performances on ImageNet remain consistent with domain shift with varying medical datasets and imaging modalities. Another question is whether pretraining helps in CXR, OCT, and WSI interpretation. 
Pretraining on datasets like ImageNet can improve the models' performance. However, performance on medical imaging tasks varies depending on several factors including data type used for training, pretrained models, and downstream medical data. Domain shift also adds variability in prediction impacting the performance globally. Improved performance can be achieved through fine-tuning with other settings. Careful fine-tuning is required when there is a difference between the pretrained data and downstream data. Transferring weights without any adaptation does not guarantee optimal performance on shifts. 

Improvement in pretraining will be pronounced in modalities where visual features are more or less complementary to those in natural images. The key to maximizing the benefits lies in careful adaptation and fine-tuning to specific medical imaging contexts. In the OCT dataset, there is a drop in models' performance with poor accuracies and AUCs. However, in the WSI dataset, the performance has been boosted significantly compared to CXR and OCT. Moreover, during pretraining with freeze layers, WSI performance has increased by 20\% compared to the other two imaging modalities. 

This observation indicates that for medical image classification, the performance of models varies significantly across modalities like CXR, OCT, and WSI. The reason behind this is that each modality provides inherent characteristics and information content. Additionally, each modality presents unique challenges that require tailored approaches in model design, with specific requirements of model parameters for training, and evaluation. The differences in structure with varying modalities are defined by anatomical regions like lungs and heart in CXR. The variability in CXRs can come from different patient positions, X-ray machines, and acquisition settings. OCT images with axial and lateral dimensions show layered structures of tissue, such as retinal layers. Variability in OCT can arise from different eye conditions, imaging angles, and devices, that affect the appearance of the retinal layers and structure. WSI with tiling contains highly detailed and complex information at the cellular level that has intricate patterns and subtle characteristics in cellular morphology. Biological variability at the cellular level is high, with different staining techniques, tumor heterogeneity, and varying tissue types. All these parameters lead to complexity or ease in model generalization. WSI images are usually high-resolution images and require tiling. These data contain highly detailed and complex information at the cellular level which requires the models to learn intricate patterns and subtle characteristics in cellular morphology. The scarcity of annotated data makes it even harder to train models. Models need to handle and integrate information from multiple tiles. Biological variability at the cellular level is high, with different staining techniques, tumor heterogeneity, and varying tissue types, leading to more complexity in model generalization. WSI datasets tend to be more limited in size compared to CXR which are widely available. The variation in the performance of deep learning models across CXR, WSI, and OCT modalities is due to differences in image resolution, information content, data availability, pre-processing needs, model architecture complexity, and clinical variability. 

Direct application of one imaging modality to another modality leads to performance degradation due to domain shift and feature mismatch. It suffers from significant challenges due to the differences in dimensionality, image characteristics, and features of specific diagnoses of each modality. The visual markers of diseases look different as this information varies across modalities, anatomical structure, and also tissue types. The learned features from CXRs may not adequately capture the complexity of OCT and WSI. Implementing any methods requires careful consideration of the specific medical imaging tasks and available datasets. 

Within the same modality, results may vary across datasets irrespective of geography. Determining the optimal model complexity for a specific medical image classification task involves carefully balancing the model’s ability to learn from the complexity of the data with its ability to generalize to new unseen data. The results demonstrate the need to carefully consider the architectural design of a model beyond just the number of parameters 
to deal with real-world scenarios. If we consider the architecture of the models without pretraining, we find that architectural choice matters. The choice of architecture of the models within the family may influence the performance across datasets of the same modality. Less parameterized models can also perform comparatively, enabling lower hardware requirements and less training time. TL through pretrained models might not be decisive for medical imaging data. The drawback of selecting larger complex models with a huge number of parameters will create restrictions as these would require higher and more efficient resources, larger training time, and prone to overfitting.

\section{Challenges And Future Prospects}\label{sec6}
Different modalities have vastly different image resolutions and sizes. Each imaging modality diagnoses different pathologies, so requires a versatile model to identify a wide range of conditions. A single model architecture that performs well across all imaging modalities poses significant challenges due to the diversity of data, its size, and the requirements that are specific to each imaging task. If the training dataset includes real-world challenges and variations similar to those of clinical datasets, the models will be efficient at handling such challenges. Conventional supervised learning is the most commonly used technique in machine learning applications. Although training in a supervised manner is an integral part of building intelligent models, the transfer of knowledge between categories is an essential part of scaling to several added categories. Using TL instead of fine-tuning the entire model, selectively fine-tune certain layers of the networks that are more likely to capture domain-specific features. TL has solved the problem of training data insufficiency and time constraints. Most TL approaches rely on creating connections in embedding or labeling spaces between the source domains and the target domains.  Domain relevance ensures that the learned features apply to the target clinical tasks, improving performance and robustness against noise and variations.  Also, the domain relevance of pre-training data helps to mitigate the effects of domain shift where source and target datasets differ significantly impacting transferability to real-world clinical applications.

Appropriate transfer of knowledge can occur only when the source distribution and the target distribution share the same specific modalities. In recent times, insufficient medical training data and disparities within the same domain distribution have emerged as two of the most significant challenges in machine learning. TL has raised increased attention recently for its robustness to training efficacy and shifting resilience. Nevertheless, these algorithms suffer from performance degradation when there occur domain shifts (natural to medical). Inductive learning as an alternative for the acquisition of knowledge has been applied in TL where inference about future instances is made on general patterns of the observed data, but the downside is that it is prone to noise with computation cost. Transductive learning is also one of the most active areas in TL. However, there persist several open challenges in TL that demand attention. Many existing TL algorithms depend on human intervention explicitly. Ideally, for any expert system, it is expected that the models learn a novel task independently by fully exploring the distribution leveraging algorithm. Furthermore, fitting the pre-experience of humans to TL models (such as expert radiologist views in clinical diagnosis) can meaningfully help models in developing new insight into the data. Also, another method of transferring knowledge with limited labeled examples can be promising. Training with a few examples reflects the development of human psychological insights, enabling generalization ability to new and novel classes.

With dynamic networks, attention mechanisms, and advanced pre-processing and optimization techniques it is feasible to develop a model that can effectively operate across different medical imaging tasks in clinical settings. Developing interactive tools allows clinicians to explore and query the model’s predictions, visualize different aspects of the decision-making process, and adjust the input to see how the predictions of models change. Identification and rectification of errors in predictions is another important concern in deep learning models. The problem of addressing and aligning with real clinical needs can be mitigated by involving clinicians in the loop during the model design and development. Clinicians can validate the model’s predictions against their expertise and ensure that the model’s reasoning aligns with established medical knowledge.

\section{Conclusion}\label{sec7}

From this study, we tried to find answers for the performances of DCNNs with TL for CXR, OCT, and WSI datasets with two initialization settings. The study analyzed the performance of various DL models when applied to different medical imaging classification tasks. The study found that within architectural families, the increase in parameters did not guarantee the highest performance. Furthermore, fine-tuning with different settings improved the performance of the models, but this may vary across datasets of the same modality. The results also showed that the performance of the models varied across different architectural families. This performance is not conclusive for interpretation in CT, MRI, USD, and mammogram images. These insights have important implications for the design and deployment of medical image classification systems. By understanding the conditions under which pretraining is most beneficial, researchers and practitioners can make more informed decisions about the use of transfer learning in their medical imaging applications. Further, exploration of these topics can help in advancing the field of medical image analysis and improve the performance of AI-powered clinical decision support systems. Expert knowledge integration i.e. incorporating expert annotations, medical atlases, or known anatomical structures into training will improve interpretability and performance.

\bibliographystyle{plain}
\bibliography{preprint}
\end{document}